\documentclass[letterpaper,english,aps,prb,twocolumn,showpacs,superscriptaddress]{revtex4}
\usepackage{amsmath}
\usepackage{graphicx}
\usepackage{amssymb}

\makeatletter
\usepackage{babel}
\makeatother

\begin{document}
\newcommand{\bra}[1]{\left\langle #1\right|}
\newcommand{\ket}[1]{\left|#1\right\rangle }
\newcommand{\braket}[2]{\left\left \langle#1|#2\right\rangle }
\newcommand{\Tr}{\mathrm{Tr}}
\newcommand{\commute}[2]{\left[#1,#2\right]}
\newcommand{\anticommute}[2]{\left\{  #1,#2\right\}  }
\newcommand{\expect}[1]{\left\langle #1\right\rangle }
\newcommand{\sans}[1]{\mathsf{#1}}

\preprint{This is a PREPRINT -- please do not distribute!!}

\title{Exponential decay in a spin bath}

\author{W. A. Coish}

\affiliation{Institute for Quantum Computing and Department of Physics and Astronomy,
University of Waterloo, 200 University Ave. W., Waterloo, ON, N2L 3G1,
Canada}

\affiliation{Department of Physics, University of Basel, Klingelbergstrasse 82,
4056 Basel, Switzerland}

\author{Jan Fischer}

\affiliation{Department of Physics, University of Basel, Klingelbergstrasse 82,
4056 Basel, Switzerland}

\author{Daniel Loss}

\affiliation{Department of Physics, University of Basel, Klingelbergstrasse 82,
4056 Basel, Switzerland}

\date{\today}

\begin{abstract}
We show that the coherence of an electron spin interacting with a
bath of nuclear spins can exhibit a well-defined purely exponential
decay for special (`narrowed') bath initial conditions in the presence
of a strong applied magnetic field. This is in contrast to the typical
case, where spin-bath dynamics have been investigated in the non-Markovian
limit, giving super-exponential or power-law decay of correlation
functions. We calculate the relevant decoherence time $T_{2}$ explicitly
for free-induction decay and find a simple expression with dependence
on bath polarization, magnetic field, the shape of the electron wave
function, dimensionality, total nuclear spin $I$, and isotopic concentration
for experimentally relevant heteronuclear spin systems.
\end{abstract}

\pacs{03.65.Yz, 72.25.Rb, 31.30.Gs }

\maketitle

\section{Introduction}

There are many proposals to use the spin states of electrons in confined
structures for coherent manipulation, leading to applications in quantum
information processing and ultimately, quantum computation.\citep{loss:1998a,hanson:2006b,leuenberger:2001a,vrijen:2000a,jelezko:2004a}
A series of recent experiments on such spin states in quantum dots,\citep{petta:2005a,koppens:2006a}
electrons bound to phosphorus donors in silicon,\citep{abe:2004a}
NV centers in diamond,\citep{jelezko:2004a,childress:2006a,hanson:2006c}
and molecular magnets\citep{ardavan:2007a,troiani:2007a} have shown
that the hyperfine interaction between confined electron spins and
nuclear spins in the surrounding material is the major obstacle to
maintaining coherence in these systems. 

Previous studies of this decoherence mechanism have pointed to the
non-Markovian nature of a slow nuclear-spin environment, leading to
non-exponential coherence decay.\citep{khaetskii:2002a,merkulov:2002a,desousa:2003a,breuer:2004a,coish:2004a,deng:2006a,witzel:2007a,yao:2007a,koppens:2007a,erlingsson:2004a,yuzbashyan:2005a,alhassanieh:2006a,chen:2007a}
These results suggest that it may be necessary to revise quantum error
correction protocols to accommodate such a `nonstandard', but ubiquitous
environment.\citep{terhal:2005a} In the present work, we show that
virtual flip-flops between electron and nuclear spins can lead to
a well-defined Markovian dynamics, giving simple exponential decay
in a large Zeeman field and for particular initial conditions (a `narrowed'
\citep{klauser:2005a} nuclear-spin state). Moreover, we calculate
the decoherence time $T_{2}$, revealing the dependence on many external
parameters for a general system. 

The rest of this paper is organized as follows: In Sec. \ref{sec:Hamiltonian}
we introduce the Hamiltonian for the Fermi contact hyperfine interaction
and derive an effective Hamiltonian for electron spin dynamics which is valid
in a strong magnetic field. In Sec. \ref{sec:Markov-approximation} we
present the Markov approximation and its range of validity, giving
an analytical expression for the decoherence time $T_{2}$. We also
give bounds for the non-Markovian corrections to our expression. Sec.
\ref{sec:Homonuclear-system} gives a discussion of the decoherence
rate for a homonuclear system and in Sec. \ref{sec:Heteronuclear-system}
we generalize these results for a heteronuclear spin bath, providing
explicit analytical expressions for $T_{2}$ within our Born-Markov
approximation. We conclude in Sec. \ref{sec:Conclusions} and present
additional technical details in Appendices \ref{sec:Continuum-limit}-\ref{sec:Decoherence-rate}.

\section{Hamiltonian\label{sec:Hamiltonian}}

We begin from the Hamiltonian for the Fermi contact hyperfine interaction
between a localized spin-1/2 $\mathbf{S}$ and an environment of nuclear
spins, 

\begin{equation}
H_{\mathrm{hf}}=bS^{z}+b\sum_{k}\gamma_{k}I_{k}^{z}+\mathbf{S}\cdot\mathbf{h};\,\,\,\,\,\mathbf{h}=\sum_{k}A_{k}\mathbf{I}_{k}.\label{eq:HFHamiltonian}\end{equation}
Here, $\mathbf{I}_{k}$ is the nuclear spin operator for the spin
at site $k$ with associated hyperfine coupling constant $A_{k}$,
$b=g^{*}\mu_{\mathrm{B}}B$ is the electron Zeeman splitting in an
applied magnetic field $B$ and $\gamma_{k}$ is the nuclear gyromagnetic
ratio in units of the electron gyromagnetic ratio (we set $\hbar=1$):
$\gamma_{k}=g_{I_{k}}\mu_{N}/g^{*}\mu_{B}$. For an electron with
envelope wave function $\psi(\mathbf{r})$, we have $A_{k}=v_{0}A^{i_{k}}|\psi(\mathbf{r}_{k})|^{2}$,
where $A^{i_{k}}$ is the total coupling constant to a nuclear spin
of species $i_{k}$ at site $k$ and $v_{0}$ is the volume of a unit
cell containing one nucleus. For convenience, we define $A=\sqrt{\sum_{i}\nu_{i}\left(A^{i}\right)^{2}}$,
where $\nu_{i}$ is the relative concentration of isotope $i$. The
envelope function $\psi(\mathbf{r})$ of the bound electron has finite
extent, and consequently there will be a finite number $\sim N$ of
nuclei with appreciable $A_{k}$. For typical quantum dots, $N\sim10^{4}-10^{6}$,
and for donor impurities or molecular magnets, $N\sim10^{2}-10^{3}$.
In Eq. (\ref{eq:HFHamiltonian}) we have neglected the anisotropic
hyperfine interaction, dipole-dipole interaction between nuclear spins,
and nuclear quadrupolar splitting, which may be present for nuclear
spin $I>1/2$. The anisotropic hyperfine interaction gives a small
correction for electrons in a primarily $s$-type conduction band,\citep{abragam:1962a}
such as in III-V semiconductors or Si. Nuclear dipole-dipole coupling
can give rise to dynamics in the spin bath, which can lead to electron-spin
decay due to spectral diffusion on a time scale found to be $T_{M}\sim10-100\,\mu\mathrm{s}$
for GaAs quantum dots.\citep{desousa:2003a,yao:2006a,witzel:2006a}
These times are one to two orders of magnitude longer than the $T_{2}$
we predict for a GaAs quantum dot carrying $N=10^{5}$ nuclei (see
Fig. \ref{fig:InGaAsDecayRates}, below). For smaller systems, we
expect the decay mechanism discussed here to dominate dipole-dipole
effects substantially. The quadrupolar splitting has also been measured
for nanostructures in GaAs, giving inverse coupling strengths on the
order of $100\,\mu\mathrm{s}$,\citep{yusa:2005a} comparable to the
dipole-dipole coupling strength, so quadrupolar effects should become
relevant on comparable time scales.

For large $b$, we divide $H_{\mathrm{hf}}=H_{0}+V_{\mathrm{ff}}$
into an unperturbed part $H_{0}$ that preserves $S^{z}$ and a term
$V_{\mathrm{ff}}=\frac{1}{2}\left(S_{+}h_{-}+S_{-}h_{+}\right)$ that
leads to energy non-conserving flip-flops between electron and nuclear
spins.\citep{coish:2004a} We eliminate $V_{\mathrm{ff}}$ to leading
order by performing a Schrieffer-Wolff-like transformation: $\overline{H}=e^{S}H_{\mathrm{hf}}e^{-S}\approx H=H_{0}+\frac{1}{2}\commute{S}{V_{\mathrm{ff}}}$,
where $S=\frac{1}{\mathsf{L_{0}}}V_{\mathrm{ff}}$, and $\mathsf{L_{0}}$
is the unperturbed Liouvillian, defined by $\mathsf{L_{0}}O=\commute{H_{0}}{O}$.
The resulting effective Hamiltonian is of the form \citep{shenvi:2005a,yao:2006a}
(see Appendix \ref{sec:Effective-Hamiltonian})

\begin{equation}
H=\left(\omega+X\right)S^{z}+D.\label{eq:HeffectiveGeneralForm}\end{equation}

The operators $\omega$ and $D$ are diagonal with respect to a product-state
basis of $I_{k}^{z}$-eigenstates $\bigotimes_{k}\ket{I_{i_{k}}m_{k}}$,
whereas the term $X$ is purely off-diagonal in this basis, leading
to correlations between different nuclei. We neglect corrections to
the diagonal part of $H$ of order $\sim A^{2}/Nb$, but retain the
term of this size in the off-diagonal part $X$. This approximation
is justified since, as we will show, the bath correlation time $\tau_{c}$
is much shorter than the time scale where these diagonal corrections
become relevant for sufficiently large Zeeman splitting $b\gg A$,
where a Born-Markov approximation is valid: $\tau_{c}\sim N/A\ll Nb/A^{2}$.
In addition, we ignore corrections to $X$ that are smaller by the
factors $A_{k}/b\sim A/Nb\ll1$ and $\gamma_{k}\sim10^{-3}$. Under
these approximations, the various terms in Eq. (\ref{eq:HeffectiveGeneralForm})
are given by (see also Appendix \ref{sec:Effective-Hamiltonian}):\begin{eqnarray}
\omega & \simeq & b+h^{z},\,\, D\simeq b\sum_{k}\gamma_{k}I_{k}^{z},\label{eq:OmegaDef}\\
X & \simeq & \frac{1}{2}\sum_{k\ne l}\frac{A_{k}A_{l}}{\omega}I_{k}^{-}I_{l}^{+}.\label{eq:XDef}\end{eqnarray}

\section{Markov approximation\label{sec:Markov-approximation}}

For large $b$, $H_{\mathrm{hf}}$ leads only to incomplete decay
of the longitudinal spin $\left\langle S_{z}\right\rangle _{t}$.\citep{coish:2004a}
However, it is still possible for the transverse spin $\left\langle S_{+}\right\rangle _{t}$
to decay fully \citep{deng:2006a} through a pure dephasing process,
which we now describe in detail. We assume that the electron and nuclear
systems are initially unentangled with each other and that the nuclear
spin system is prepared in a narrowed state (an eigenstate of the
operator $\omega$: $\omega\ket{n}=\omega_{n}\ket{n}$) through a
sequence of weak measurements,\citep{klauser:2005a,giedke:2006a,stepanenko:2005a}
polarization pumping,\citep{ramon:2007a} frequency focusing under
pulsed optical excitation,\citep{greilich:2007a} or by any other
means. For these initial conditions, dynamics of the transverse electron
spin $\left\langle S_{+}\right\rangle _{t}$ are described by the
exact equation of motion:\citep{coish:2004a}\begin{eqnarray}
\dot{\left\langle S_{+}\right\rangle }_{t} & = & i\omega_{n}\left\langle S_{+}\right\rangle _{t}-i\int_{0}^{t}dt^{\prime}\Sigma(t-t^{\prime})\left\langle S_{+}\right\rangle _{t^{\prime}},\label{eq:SplusGME}\\
\Sigma(t) & = & -i\mathrm{Tr}S_{+}\mathsf{L}e^{-i\mathsf{Q}\mathsf{L}t}\mathsf{Q}\mathsf{L}\ket{n}\bra{n}S_{-}.\label{eq:SelfEnergyDefinition}\end{eqnarray}
Here, $\mathsf{L}$ and $\mathsf{Q}$ are superoperators, defined
by their action on an arbitrary operator $O$: $\mathsf{L}O=\commute{H}{O}$,
$\mathsf{Q}O=\left(1-\ket{n}\bra{n}\mathrm{Tr}_{I}\right)O$, where
$\mathrm{Tr}_{I}$ indicates a partial trace over the nuclear spin
system. 

To remove fast oscillations in $\left<S_{+}\right>_{t}$ we transform
to a rotating frame, in which we define the coherence factor $x_{t}=2\exp\left[-i(\omega_{n}+\Delta\omega)t\right]\left\langle S_{+}\right\rangle _{t}$
and associated memory kernel $\tilde{\Sigma}(t)=\exp\left[-i(\omega_{n}+\Delta\omega)t\right]\Sigma(t)$,
with frequency shift determined self-consistently through $\Delta\omega=-\mathrm{Re}\int_{0}^{\infty}dt\tilde{\Sigma}(t)$.
Additionally, we change integration variables to $\tau=t-t^{\prime}$.
The equation of motion for $x_{t}$ then reads \begin{equation}
\dot{x}_{t}=-i\int_{0}^{t}d\tau\tilde{\Sigma}(\tau)x_{t-\tau}.\label{eq:EOMforx}\end{equation}
If $\tilde{\Sigma}(\tau)$ decays to zero sufficiently quickly %
\footnote{\label{foot:NoMarkov}The integral in Eq. (\ref{eq:SplusExponentialSolution})
becomes undefined if the memory kernel has an asymptotic time dependence
$\tilde{\Sigma}(t)\sim1/t^{\alpha}$, where $\alpha\le1$, and consequently
the Markov approximation breaks down in this case. A weaker version
of Markovian violation can occur more generally for $\alpha\le2$,
in which case the bound, Eq. (\ref{eq:NonMarkovianBound}), may still
be small for times $t\sim T_{2}$, but grows unbounded in time. This
situation occurs, for example, in the ohmic spin-boson model.\citep{divincenzo:2005a}%
} on the time scale $\tau_{c}\ll T_{2}$, where $T_{2}$ is the decay
time of $x_{t}$, we can approximate $x_{t-\tau}\approx x_{t}$ and
extend the upper limit on the integral to $t\to\infty$ (Markov approximation),
giving an exponential coherence decay with a small error $\epsilon(t)$:\begin{equation}
x_{t}=\exp\left(-t/T_{2}\right)x_{0}+\epsilon(t),\,\,\,\frac{1}{T_{2}}=-\mathrm{Im}\int_{0}^{\infty}dt\tilde{\Sigma}(t).\label{eq:SplusExponentialSolution}\end{equation}

\begin{figure}
\includegraphics[scale=0.8]{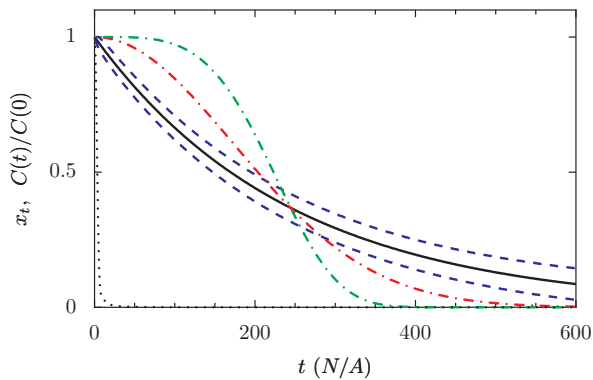}

\caption{\label{fig:ExponentialDecayWithBounds}(Color online) Exponential
decay $x_{t}=\exp\left(-t/T_{2}\right)$ (solid line) and maximum
error bounds $x_{t}\pm\left|\epsilon(t)\right|_{\mathrm{max}}$ (dashed
lines), found by numerical integration of Eq. (\ref{eq:NonMarkovianBound})
with parameters for a two-dimensional quantum dot (before Eq. (\ref{eq:HomonuclearFreeInductionHighField})),
$I=3/2$ and $A/b=1/20$. For comparison, we show the decay curves
for super-exponential forms $\exp\left[-\left(t/T_{2}\right)^{2}\right]$
and $\exp\left[-\left(t/T_{2}\right)^{4}\right]$ (dot-dashed lines)
and rapidly decaying bath correlation function $C(t)/C(0)$ (dotted
line, see Eqs. (\ref{eq:HomonuclearFreeInductionRate}) and (\ref{eq:XXCorrelator})).}
\end{figure}

The non-Markovian correction $\epsilon(t)$ can be bounded precisely
if $\tilde{\Sigma}(t)$ is known:\citep{fick_markov_bound:1990a}\begin{equation}
\left|\epsilon(t)\right|\le\left|\epsilon(t)\right|_{\mathrm{max}}=2\int_{0}^{t}dt^{\prime}\left|\int_{t^{\prime}}^{\infty}dt^{\prime\prime}\tilde{\Sigma}(t^{\prime\prime})\right|.\label{eq:NonMarkovianBound}\end{equation}
Eq. (\ref{eq:NonMarkovianBound}) gives a hard bound on the validity
of the Markov approximation, and consequently, any corrections to
the exponential decay formula. Fig. \ref{fig:ExponentialDecayWithBounds}
demonstrates an application of Eqs. (\ref{eq:SplusExponentialSolution})
and (\ref{eq:NonMarkovianBound}) for decay in a homonuclear spin
system, which we discuss below.%

\section{Homonuclear\emph{ }system\label{sec:Homonuclear-system}}

If only one spin-carrying nuclear isotope is present, $\gamma_{k}=\gamma$,
independent of the nuclear site. We then approximate $\Sigma(t)$
to leading order in the perturbation $V=XS^{z}$ (Born approximation,
see Appendix \ref{sec:Born-approximation}) by expanding Eq. (\ref{eq:SelfEnergyDefinition})
through iteration of the Dyson identity: $e^{-i\mathsf{LQ}t}=e^{-i\mathsf{L}_{0}\mathsf{Q}t}-i\int_{0}^{t}dt^{\prime}e^{-i\mathsf{L}_{0}\mathsf{Q}(t-t^{\prime})}\mathsf{L}_{V}\mathsf{Q}e^{-i\mathsf{LQ}t^{\prime}}$,
where $\mathsf{L}_{V}O=\commute{V}{O}$. Higher-order corrections
to the Born approximation will be suppressed by the small parameter
$A/\omega_{n}$.\citep{coish:2004a} Inserting the result into Eq.
(\ref{eq:SplusExponentialSolution}) we find:\begin{equation}
\frac{1}{T_{2}}=\mathrm{Re}\int_{0}^{\infty}dte^{-i\Delta\omega t}\left\langle X(t)X\right\rangle ;\, X(t)=e^{-i\omega t}Xe^{i\omega t}.\label{eq:HomonuclearFreeInductionRate}\end{equation}
Here, $\left<\cdots\right>=\bra{n}\cdots\ket{n}$ denotes an expectation
value with respect to the initial nuclear state. Eq. (\ref{eq:HomonuclearFreeInductionRate})
resembles the standard result for pure dephasing in a weak coupling
expansion, where $X(t)$ would represent the bath operator in the
interaction picture with an independent bath Hamiltonian. However,
for the spin bath there is no such weak coupling expansion, and $X(t)$
appears in the interaction picture with $\omega$, the same operator
that provides an effective level splitting for the system. Additionally,
the general result for a heteronuclear system including inter-species
flip-flops cannot be written in such a compact form.\citep{coish:2007a}

Previously, it has been shown that a Born-Markov approximation to
\emph{second} order in $V_{\mathrm{ff}}$ leads to no decay.\citep{coish:2004a}
In contrast, a Born-Markov approximation applied to the effective
Hamiltonian leads directly to a result that is \emph{fourth} order
in $V_{\mathrm{ff}}$ {[}Eq. (\ref{eq:HomonuclearFreeInductionRate})],
describing dynamics that become important at times longer than the
second-order result. It is not \emph{a priori} obvious that the effective
Hamiltonian, evaluated only to second order in $V_{\mathrm{ff}}$,
can be used to accurately calculate rates to fourth order in $V_{\mathrm{ff}}$.
We have, however, verified that all results we present here are equivalent
to a direct calculation expanded to fourth order in $V_{\mathrm{ff}}$
at leading order in $A/b\ll1$.\citep{coish:2007a}

If the initial nuclear polarization is smooth on the scale of the
electron wave function, the matrix elements of operators like $I_{k}^{\pm}I_{k}^{\mp}$
can be replaced by average values. Neglecting corrections that are
small in $A/Nb\ll1$, this gives (see also Appendix \ref{sec:Decoherence-rate}):\begin{equation}
C(t)=\left<X(t)X(0)\right>=\frac{c_{+}c_{-}}{4\omega_{n}^{2}}\sum_{k\ne l}A_{k}^{2}A_{l}^{2}e^{-i(A_{k}-A_{l})t}.\label{eq:XXCorrelator}\end{equation}
Above, we have introduced the coefficients $c_{\pm}=I(I+1)-\left<\left<m(m\pm1)\right>\right>$
and the double angle bracket indicates an average over $I_{k}^{z}$
eigenvalues $m$.\citep{coish:2004a} 

In the limit $N\gg1$ we can include the term $k=l$ in Eq. (\ref{eq:XXCorrelator})
and perform the continuum limit $\Sigma_{k}\to\int dk$ with small
corrections. For an isotropic electron wave function of the form $\psi(r)=\psi(0)e^{-\left({r/r}_{0}\right)^{q}/2}$
containing $N$ nuclei within radius $r_{0}$ in $d$ dimensions,
the hyperfine coupling constants are distributed according to $A_{k}=A_{0}\exp\left[-\left(k/N\right)^{q/d}\right]$,
where $k$ is a non-negative index, and we choose $A_{0}$ to normalize
$A_{k}$ according to $A=\int_{0}^{\infty}dkA_{k}$\citep{coish:2004a}
(see also Appendix \ref{sec:Continuum-limit}).

\begin{figure}
\includegraphics[clip,scale=0.8]{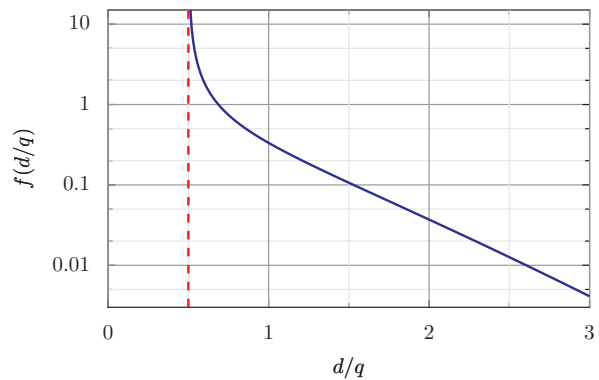}\vspace{-0.3cm}

\caption{\label{fig:Geometrical-factor} (Color online) Geometrical factor
$f(d/q)$ from Eq. (\ref{eq:ScalingFunction}), where $d=1,2,3$ is
the dimension and $q$ characterizes the electron envelope function
$\psi(r)=\psi(0)\exp\left[-\left(r/r_{0}\right)^{q}/2\right]$.\vspace{-0.3cm}}

\end{figure}

After performing the continuum limit, $C(t)$ will decay, with characteristic
time $\tau_{c}$ given by the inverse bandwidth of nuclear flip-flop
excitations $\tau_{c}\sim1/A_{0}\sim N/A$. For large $b$, $1/T_{2}$
will be suppressed due to the smallness of $X$ (see Eq. (\ref{eq:XDef})),
whereas $\tau_{c}$ remains fixed. At sufficiently large $b$, it
will therefore be possible to reach the Markovian regime, where $\tau_{c}$
is short compared to $T_{2}$: $\tau_{c}/T_{2}\ll1$. Evaluating the
time integral in Eq. (\ref{eq:HomonuclearFreeInductionRate}), we
find the general result to leading order in $A/\omega_{n}$ (see Appendix
\ref{sec:Decoherence-rate}):\begin{eqnarray}
\frac{1}{T_{2}} & = & \frac{\pi}{4}c_{+}c_{-}f\left(\frac{d}{q}\right)\left(\frac{A}{\omega_{n}}\right)^{2}\frac{A}{N},\label{eq:GeneralDecayRate}\\
f(r) & = & \frac{1}{r}\left(\frac{1}{3}\right)^{2r-1}\frac{\Gamma(2r-1)}{\left[\Gamma(r)\right]^{3}},\,\,\, r>1/2.\label{eq:ScalingFunction}\end{eqnarray}
In Eq. (\ref{eq:GeneralDecayRate}), $A/N$ sets the scale for the
maximum decay rate in the perturbative regime, the coefficients $c_{\pm}$
set the dependence on the initial nuclear polarization $p$ (\emph{e.g.},
with $I=1/2$, we have $c_{+}c_{-}=(1-p^{2})/4$), $A/\omega_{n}<1$
gives the small parameter which controls the Born approximation, and
$f(d/q)$ is a geometrical factor (plotted in Fig. \ref{fig:Geometrical-factor}).
$f(d/q)$ is exponentially suppressed for $d/q>1$ ($f(r)\propto(1/3)^{2r-1}(1/r)^{r},\, r>1$),
but $f(d/q)\to\infty$ for $d/q-1/2\to0^{+}$. Due to this divergence,
no Markov approximation is possible (within the Born approximation)
for $d/q\le1/2$. We understand the divergence in $f(d/q)$ explicitly
from the asymptotic dependence of $C(t)$ at long times: $C(t)\propto1/t^{2d/q},\, t\gg N/A,$~$d/q<2$.\footnotemark[\value{footnote}]
Surprisingly, there is a difference of nearly \emph{two orders} of
magnitude in $1/T_{2}$ going from a two-dimensional (2D) quantum
dot with Gaussian envelope function ($d=2$, $q=2$, $d/q$=1) to
a donor impurity with a hydrogen-like exponential wave function ($d=3,\, q=1,\, d/q=3$),
if all other parameters are fixed (see Fig. \ref{fig:Geometrical-factor}).

We now specialize to an initial uniform unpolarized spin bath, which
is nevertheless narrowed: $\omega\ket{n}=b\ket{n}$, with equal populations
of all nuclear Zeeman levels (i.e., $\left<\left<m\right>\right>=0$
and $\left<\left<m^{2}\right>\right>=\frac{1}{3}I(I+1)$). For a 2D
quantum dot with a Gaussian envelope function ($d=q=2$) we find,
from Eqs. (\ref{eq:GeneralDecayRate}) and (\ref{eq:ScalingFunction}):\begin{equation}
\frac{1}{T_{2}}=\frac{\pi}{3}\left(\frac{I(I+1)A}{3b}\right)^{2}\frac{A}{N}.\label{eq:HomonuclearFreeInductionHighField}\end{equation}

There are two remarkable features of this surprisingly simple result.
First, the condition for the validity of the Markov approximation,
$T_{2}>\tau_{c}\sim N/A$ will be satisfied whenever $A/b<1$, which
is the same condition that validates a Born approximation. Second,
$1/T_{2}$ has a very strong dependence on the nuclear spin ($1/T_{2}\propto I^{4}$).
Thus, systems with large-spin nuclei such as In ($I_{\mathrm{In}}=9/2$)
will show relatively significantly faster decay (see, \emph{e.g.},
Fig. \ref{fig:InGaAsDecayRates}).

\section{Heteronuclear system\label{sec:Heteronuclear-system}}

\begin{figure}
\includegraphics[scale=0.8]{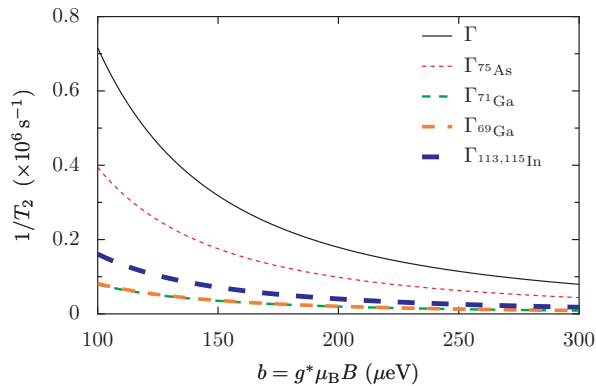}\vspace{-0.3cm}

\caption{\label{fig:InGaAsDecayRates}(Color online) Decay rates for an $\mathrm{In}_{x}\mathrm{Ga}_{1-x}\mathrm{As}$
quantum dot with In doping $x=0.05$. Here, we have assumed $N=10^{5}$
and used values of $\nu_{i}$ and $A^{i}$ for GaAs from Ref. {[}\onlinecite{paget:1977a}]:
$A^{^{75}\mathrm{As}}=86\,\mu eV,\, A^{^{69}\mathrm{Ga}}=74\,\mu eV,\, A^{^{71}\mathrm{Ga}}=96\,\mu eV,$~$\nu_{^{75}\mathrm{As}}=0.5,\,\nu_{^{69}\mathrm{Ga}}=0.3(1-x),{\,\nu}_{^{71}\mathrm{Ga}}=0.2(1-x)$.
The hyperfine coupling for In in InAs was taken from Ref. {[}\onlinecite{liu:2007a}]:
$A^{^{113}\mathrm{In}}\approx A^{^{115}\mathrm{In}}\approx A^{\mathrm{In}}=170\,\mu eV$,
$\nu_{\mathrm{In}}=x/2$.\vspace{-0.3cm}}

\end{figure}
For sufficiently large $b$ ($\left|\gamma_{k}-\gamma_{k^{\prime}}\right|b\gg\left|A_{k}-A_{k^{\prime}}\right|\sim A/N$),
heteronuclear flip-flops between two isotopic species with different
$\gamma_{k}$ are forbidden due to energy conservation. In this case,
$1/T_{2}$ is given in terms of an incoherent sum, $1/T_{2}=\Gamma=\sum_{i}\Gamma_{i}$,
where $\Gamma_{i}$ is the contribution from flip-flops between nuclei
of the common species $i$. Assuming a uniform distribution of all
isotopes in a 2D quantum dot with a Gaussian envelope function, we
find (see also Appendix \ref{sec:Decoherence-rate}):\begin{equation}
\Gamma_{i}=\frac{1}{T_{2}^{i}}=\nu_{i}^{2}\frac{\pi}{3}\left(\frac{I_{i}(I_{i}+1)A^{i}}{3b}\right)^{2}\frac{A^{i}}{N}.\label{eq:HeteronuclearRate}\end{equation}
The quadratic dependence on isotopic concentration $\nu_{i}$ is particularly
striking. Due to this dependence, electron spins in GaAs, where Ga
has two naturally occurring isotopic species, whereas As has only
one, will show a decay predominantly due to flip-flops between As
spins. This is in spite of the fact that all isotopes in GaAs have
the same nuclear spin and nominally similar hyperfine coupling constants
(see Fig. \ref{fig:InGaAsDecayRates}). Interestingly, we note that
the relatively large flip-flop rates for In and As, due to large nuclear
spin and isotopic concentration, respectively, may partly explain
why only Ga (and not In or As) spins have been seen to contribute
to coherent effects in experiments on electron transport through (In/Ga)As
quantum dots.\citep{ono:2004a} The same effect may also explain why
polarization appears to be transferred more efficiently from electrons
to As (rather than Ga) in GaAs quantum dots.\citep{yacoby:2007a}

\section{Conclusions\label{sec:Conclusions}}

We have shown that a single electron spin can exhibit a purely
exponential decay for narrowed nuclear-spin bath initial conditions
and in the presence of a sufficiently large electron Zeeman splitting
$b$. This work may be important for implementing existing quantum
error correction schemes, which typically assume exponential decay of
correlation functions due to a Markovian environment. In the limit of
large Zeeman splitting $b>A$, where a Born-Markov approximation is
valid, we have found explicit analytical expressions for the
decoherence time $T_{2}$, giving explicit dependences on the electron
wave function, magnetic field, bath polarization, nuclear spin, and
isotopic abundance for a general nuclear spin bath.  Moreover, within
the Born-Markov approximation, we have found a divergence in the
decoherence rate $1/T_2$ for a one-dimensional quantum dot, indicating
a breakdown of the Markov approximation in this case.

\begin{acknowledgments}
We thank J. Baugh, D. Klauser, and F. H. L. Koppens for useful discussions.
We acknowledge funding from the Swiss NSF, NCCR Nanoscience, JST ICORP,
QuantumWorks, and an Ontario PDF (WAC).
\end{acknowledgments}
\appendix

\section{Continuum limit\label{sec:Continuum-limit}}

In this appendix, we describe how the dimensionality $d$ and envelope
wave function shape parameter $q$ are defined. For further details
on the definition of these quantities, see Ref. {[}\onlinecite{coish:2004a}].
For a homonuclear spin system, the hyperfine coupling constants are
given by\begin{equation}
A_{k}=Av_{0}\left|\psi(\mathbf{r}_{k})\right|^{2},\label{eq:AkDefinition}\end{equation}
 where $A$ is the total hyperfine coupling constant, $v_{0}$ is
the volume occupied by a single-nucleus unit cell, and $\psi(\mathbf{r})$
is the electron envelope wave function. We assume an isotropic electron
envelope:\begin{equation}
\psi(r_{k})=\psi(0)e^{-\frac{1}{2}\left(\frac{r_{k}}{r_{0}}\right)^{q}},\label{eq:EnvelopeFunction}\end{equation}
 where $r_{0}$ is the effective Bohr radius, defined as the radial distance
enclosing $N$ nuclear spins, and $r_{k}$ is the radial distance
enclosing $k$ spins. In $d$ dimensions:\begin{equation}
\frac{\mathrm{vol}(k\,\mathrm{spins})}{\mathrm{vol}(N\,\mathrm{spins})}=\frac{v_{0}k}{v_{0}N}=\left(\frac{r_{k}}{a_{B}}\right)^{d}.\label{eq:kdivNDefinition}\end{equation}
 Inserting Eqs. (\ref{eq:kdivNDefinition}) and (\ref{eq:EnvelopeFunction})
into Eq. (\ref{eq:AkDefinition}):\begin{equation}
A_{k}=A_{0}e^{-\left(\frac{k}{N}\right)^{q/d}}.\label{eq:Ak2}\end{equation}
To determine the $k=0$ coupling $A_{0}$, we enforce the normalization:\begin{equation}
\sum_{k}A_{k}=Av_{0}\sum_{k}\left|\psi(r_{k})\right|^{2}\approx A\int d^{3}r\left|\psi(r)\right|^{2}=A.\end{equation}
 This gives\begin{equation}
A=A_{0}\int_{0}^{\infty}dke^{-\left(\frac{k}{N}\right)^{q/d}}.\label{eq:ContinuumNormalization}\end{equation}
 Making the change of variables $u=\left(\frac{k}{N}\right)^{q/d}$,
we find immediately\begin{equation}
A=A_{0}\frac{d}{q}N\int_{0}^{\infty}duu^{\frac{d}{q}-1}e^{-u}=A_{0}N\frac{d}{q}\Gamma\left(\frac{d}{q}\right),\end{equation}
which gives the final form for $A_{k}$:\begin{equation}
A_{k}=\frac{A}{N\frac{d}{q}\Gamma\left(\frac{d}{q}\right)}e^{-\left(\frac{k}{N}\right)^{q/d}}.\label{eq:AkContLimit}\end{equation}

\section{Effective Hamiltonian\label{sec:Effective-Hamiltonian}}

In this appendix we give details leading to the derivation of the effective
Hamiltonian, described by Eqs. (2), (3), and (4) of the main text.
Similar effective Hamiltonians have been derived previously in Refs.
{[}\onlinecite{shenvi:2005a}] and {[}\onlinecite{yao:2006a}], but
due to some differences in method and approximation, we give additional
details here for the interested reader. We begin from the hyperfine
Hamiltonian:\begin{eqnarray}
H_{\mathrm{hf}} & = & H_{0}+V_{\mathrm{ff}},\label{eq:HhfDefinition}\\
H_{0} & = & \left(b+h^{z}\right)S^{z}+b\sum_{k}\gamma_{k}I_{k}^{z},\label{eq:H0Definition}\\
V_{\mathrm{ff}} & = & \frac{1}{2}\left(S_{+}h_{-}+S_{-}h_{+}\right),\label{eq:VffDefinition}\\
\mathbf{h} & = & \sum_{k}A_{k}\mathbf{I}_{k}.\label{eq:Nuclearh}\end{eqnarray}
 To find an effective Hamiltonian that eliminates the flip-flop term
$V_{\mathrm{ff}}$ at leading order, we apply a unitary transformation:\begin{equation}
\overline{H}=e^{S}H_{\mathrm{hf}}e^{-S},\label{eq:HUnitary}\end{equation}
where $S=-S^{\dagger}$ to ensure unitarity. We now expand Eq. (\ref{eq:HUnitary})
in powers of $S$, retaining terms up to $\mathcal{O}\left(V_{\mathrm{ff}}^{3}\right)$,
assuming $S\sim\mathcal{O}\left(V_{\mathrm{ff}}\right)$:\begin{multline}
\overline{H}=H_{0}+V_{\mathrm{ff}}-\commute{H_{0}}{S}-\commute{V_{\mathrm{ff}}}{S}\\
+\frac{1}{2}\commute{S}{\commute{S}{H_{0}}}+\mathcal{O}\left(V_{\mathrm{ff}}^{3}\right).\label{eq:HTransformed}\end{multline}
 To eliminate $V_{\mathrm{ff}}$ at leading order, we must choose
$S$ to satisfy $V_{\mathrm{ff}}-\commute{H_{0}}{S}=0$. The $S$
that satisfies this relation is given by\begin{equation}
S=\frac{1}{\mathsf{L}_{0}}V_{\mathrm{ff}};\;\mathsf{L}_{0}O=\commute{H_{0}}{O},\label{eq:TransformationS}\end{equation}
which is of order $V_{\mathrm{ff}}$, justifying our previous assumption:
$S\sim\mathcal{O}\left(V_{\mathrm{ff}}\right)$. Re-inserting Eq.
(\ref{eq:TransformationS}) into Eq. (\ref{eq:HTransformed}), we
find, up to corrections that are third-, or higher-order in $V_{\mathrm{ff}}$:\begin{eqnarray}
\overline{H} & = & H+\mathcal{O}\left(V_{\mathrm{ff}}^{3}\right),\label{eq:HTransformedV3Corrections}\\
H & = & H_{0}+\frac{1}{2}\commute{S}{V_{\mathrm{ff}}}.\label{eq:EffectiveHDefinition}\end{eqnarray}
 Directly evaluating Eq. (\ref{eq:TransformationS}) with $H_{0}$
defined in Eq. (\ref{eq:H0Definition}) and $V_{\mathrm{ff}}$ defined
in Eq. (\ref{eq:VffDefinition}) gives\begin{multline}
S=\frac{1}{2}\sum_{k}A_{k}\left(\frac{1}{b+h^{z}+\frac{A_{k}}{2}-b\gamma_{k}}S^{+}I_{k}^{-}\right.\label{eq:SDefinition}\\
\left.-\frac{1}{b+h^{z}-\frac{A_{k}}{2}-b\gamma_{k}}S^{-}I_{k}^{+}\right).\end{multline}
Inserting Eq. (\ref{eq:SDefinition}) into Eq. (\ref{eq:EffectiveHDefinition})
gives\begin{eqnarray}
H & = & \ket{\uparrow}\bra{\uparrow}H_{\uparrow}+\ket{\downarrow}\bra{\downarrow}H_{\downarrow},\\
H_{\uparrow} & = & \frac{1}{2}\left(b+h^{z}\right)+b\sum_{k}\gamma_{k}I_{k}^{z}+h_{\uparrow},\\
H_{\downarrow} & = & -\frac{1}{2}\left(b+h^{z}\right)+b\sum_{k}\gamma_{k}I_{k}^{z}-h_{\downarrow}.\end{eqnarray}
Here, the contributions resulting from the term second-order in $V_{\mathrm{ff}}$
are given explicitly by

\begin{multline}
h_{\uparrow}=\frac{1}{8}\sum_{kl}A_{k}A_{l}\left(\frac{1}{b+h^{z}+A_{k}/2-b\gamma_{k}}I_{k}^{-}I_{l}^{+}\right.\\
\left.+I_{l}^{-}\frac{1}{b+h^{z}-A_{k}/2-b\gamma_{k}}I_{k}^{+}\right),\label{eq:hupdefinition}\end{multline}

\begin{multline}
h_{\downarrow}=\frac{1}{8}\sum_{kl}A_{k}A_{l}\left(\frac{1}{b+h^{z}-A_{k}/2-b\gamma_{k}}I_{k}^{+}I_{l}^{-}\right.\\
+\left.I_{l}^{+}\frac{1}{b+h^{z}+A_{k}/2-b\gamma_{k}}I_{k}^{-}\right).\label{eq:hdndefinition}\end{multline}

We can rewrite $H$ in terms of spin operators using $\ket{\uparrow}\bra{\uparrow}=\frac{1}{2}+S^{z}$
and $\ket{\downarrow}\bra{\downarrow}=\frac{1}{2}-S^{z}$, which gives
Eq. (2) from the main text:\begin{eqnarray}
H & = & \left(\omega+X\right)S^{z}+D,\label{eq:HeffFinalForm}\\
X & = & \left(1-\mathsf{P}_{\mathrm{d}}\right)\left(h_{\uparrow}+h_{\downarrow}\right),\label{eq:Xeff}\\
D & = & b\sum_{k}\gamma_{k}I_{k}^{z}+\frac{1}{2}\left(h_{\uparrow}-h_{\downarrow}\right),\label{eq:Deff}\\
\omega & = & b+h^{z}+\mathsf{P}_{\mathrm{d}}\left(h_{\uparrow}+h_{\downarrow}\right).\label{eq:omegaEff}\end{eqnarray}
 In the above expressions, we have introduced the diagonal projection
superoperator $\mathsf{P}_{\mathrm{d}}O=\sum_{l}\ket{l}\bra{l}\bra{l}O\ket{l}$,
where the index $l$ runs over all nuclear-spin product states $\ket{l}=\bigotimes_{k}\ket{I_{k}m_{k}^{l}}$.
We now apply the commutation relation $\commute{I_{k}^{+}}{I_{l}^{-}}=2I_{k}^{z}\delta_{kl}$
and expand the prefactors in Eqs. (\ref{eq:hupdefinition}) and (\ref{eq:hdndefinition})
in terms of the smallness parameter $\frac{A_{k}}{b+h^{z}-b\gamma_{k}}\sim\frac{1}{N}\frac{A}{b}\ll1$.
At leading order in the expansion, we find $h_{\uparrow,\downarrow}\approx h_{\uparrow,\downarrow}^{(0)}$,
where\begin{eqnarray}
h_{\uparrow}^{(0)} & = & \frac{1}{8}\sum_{kl}\frac{A_{k}A_{l}}{b+h^{z}-b\gamma_{k}}\left(I_{k}^{-}I_{l}^{+}+I_{l}^{-}I_{k}^{+}\right),\label{eq:hupapprox1}\\
h_{\downarrow}^{(0)} & = & \frac{1}{8}\sum_{kl}\frac{A_{k}A_{l}}{b+h^{z}-b\gamma_{k}}\left(I_{k}^{+}I_{l}^{-}+I_{l}^{+}I_{k}^{-}\right).\label{eq:hdnapprox1}\end{eqnarray}
 By commuting the nuclear spin operators, Eqs. (\ref{eq:hupapprox1})
and (\ref{eq:hdnapprox1}) can be rewritten to give\begin{equation}
h_{\downarrow}^{(0)}=h_{\uparrow}^{(0)}+\frac{1}{2}\sum_{k}\frac{A_{k}^{2}}{b+h^{z}-b\gamma_{k}}I_{k}^{z}.\end{equation}
 This relation allows us to approximate the various terms in Eqs.
(\ref{eq:Xeff}), (\ref{eq:Deff}), and (\ref{eq:omegaEff}):

\begin{eqnarray}
X & \approx & \left(1-\mathsf{P}_{\mathrm{d}}\right)\left(2h_{\uparrow}^{(0)}\right),\nonumber \\
 & = & \frac{1}{4}\sum_{k\ne l}\frac{A_{k}A_{l}}{b+h^{z}-b\gamma_{k}}\left(I_{k}^{-}I_{l}^{+}+I_{l}^{-}I_{k}^{+}\right),\label{eq:Xapprox1}\end{eqnarray}
and

\begin{equation}
D\approx\sum_{k}\left(b\gamma_{k}-\frac{A_{k}^{2}}{4\left(b+h^{z}-b\gamma_{k}\right)}\right)I_{k}^{z},\label{eq:Dapprox1}\end{equation}
\begin{multline}
\omega\approx b+h^{z}\\
+\mathsf{P}_{\mathrm{d}}\left(2h_{\uparrow}^{(0)}\right)+\frac{1}{2}\sum_{k}\frac{A_{k}^{2}}{b+h^{z}-b\gamma_{k}}I_{k}^{z},\end{multline}
 or

\begin{multline}
\omega\approx b+h^{z}\\
+\frac{1}{2}\sum_{k}\frac{A_{k}^{2}}{b+h^{z}-b\gamma_{k}}\left(I_{k}(I_{k}+1)-\left(I_{k}^{z}\right)^{2}\right).\label{eq:omegaapprox1}\end{multline}
Neglecting further corrections that are smaller by the factor $b\gamma_{k}/\omega\sim\gamma_{k}\sim10^{-3}$
in Eq. (\ref{eq:Xapprox1}) and terms of order $\lesssim\sum_{k}\frac{A_{k}^{2}}{b+h^{z}-b\gamma_{k}}\sim\frac{A^{2}}{Nb}$
in Eqs. (\ref{eq:Dapprox1}) and (\ref{eq:omegaapprox1}), we arrive
immediately at Eqs. (3) and (4) of the main text. The terms of order
$\sim A^{2}/Nb$ may become important on a time scale $\tau\sim Nb/A^{2}$.
In our treatment, this time scale is long compared to the bath correlation
time $\tau_{c}\sim N/A$ in the perturbative regime $A/b<1$, and
so neglecting these terms is justified.

\section{Born approximation\label{sec:Born-approximation}}

In this appendix we give further detail on the Born approximation.
We begin from the equation of motion for the transverse spin in the
rotating frame $x_{t}$ after applying the Markov approximation, neglecting
the correction $\epsilon(t)$ (following Eq. (7)):\begin{eqnarray}
\dot{x}_{t} & = & -i\int_{0}^{\infty}d\tau\tilde{\Sigma}(\tau)x_{t},\label{eq:XDotEquation}\\
\tilde{\Sigma}(t) & = & e^{-i\left(\omega_{n}+\Delta\omega\right)t}\Sigma(t),\label{eq:SigmaTilde}\\
\Sigma(t) & = & -i\mathrm{Tr}S_{+}\mathsf{L}\mathsf{Q}e^{-i\mathsf{L}\mathsf{Q}t}\mathsf{L}\mathsf{Q}\ket{n}\bra{n}S_{-}.\label{eq:SigmaDefinition}\end{eqnarray}
 In general, it is not simple to find the exact form of the self energy
(memory kernel) $\Sigma(t)$. Fortunately, it is possible to generate
a systematic expansion in the perturbation $V=XS^{z}\propto1/b$,
valid for sufficiently large Zeeman splitting $b>A$:\citep{coish:2004a}
\begin{equation}
\Sigma(t)=\Sigma^{(2)}(t)+\Sigma^{(4)}(t)+\cdots,\end{equation}
where $\Sigma^{(n)}(t)$ indicates a term of order $\sim\mathcal{O}(V^{n})\sim\mathcal{O}\left[\left(\frac{A}{b}\right)^{n}\right]$.
The expansion is performed most conveniently in terms of the Laplace-transformed
variable\begin{equation}
\Sigma(s)=\mathcal{L}\left[\Sigma(t)\right]=\int_{0}^{\infty}dte^{-st}\Sigma(t).\label{eq:LaplaceTransform}\end{equation}
We expand the propagator $\mathcal{L}\left[e^{-i\mathsf{L}\mathsf{Q}t}\right]=\frac{1}{s+i\mathsf{L}\mathsf{Q}}$
by dividing the full Liouvillian into unperturbed and perturbed parts:
$\mathsf{L}=\mathsf{L}_{0}+\mathsf{L}_{V}$, where $\mathsf{L}_{0}$
and $\mathsf{L}_{V}$ are defined by their action on an arbitrary
operator $O$ through $\mathsf{L}_{0}O=\commute{H_{0}}{O}$ and $\mathsf{L}_{V}O=\commute{V}{O}$.
To obtain an expansion in terms of the perturbation $\mathsf{L}_{V}$,
we now iterate the Dyson identity in Laplace space:\begin{equation}
\frac{1}{s+i\mathsf{L}\mathsf{Q}}=\frac{1}{s+i\mathsf{L}_{0}\mathsf{Q}}-i\frac{1}{s+i\mathsf{L}_{0}\mathsf{Q}}\mathsf{L}_{V}\mathsf{Q}\frac{1}{s+i\mathsf{L}_{0}\mathsf{Q}}+\mathcal{O}\left(\mathsf{L}_{V}^{2}\right).\label{eq:DysonIdentity}\end{equation}
Inserting the iterated expression (Eq. (\ref{eq:DysonIdentity}))
into the Laplace-transformed version of Eq. (\ref{eq:SigmaDefinition}),
we find the self energy in Born approximation (to second order in
$V$) is

\begin{multline}
\Sigma^{(2)}(s)=-i\mathrm{Tr}\left[S_{+}\left(1-i\mathsf{L}_{0}\mathsf{Q}\frac{1}{s+i\mathsf{L}_{0}}\right)\right.\\
\left.\times\mathsf{L}_{V}\frac{1}{s+i\mathsf{L}_{0}}\mathsf{L}_{V}\ket{n}\bra{n}S_{-}\right].\end{multline}
We have simplified the above expression using the following identities
for the projection superoperators $\mathsf{Q}=\mathsf{1}-\ket{n}\bra{n}\mathrm{Tr}_{I}$
and $\mathsf{P}=1-\mathsf{Q}$:\begin{eqnarray}
\mathsf{P}\mathsf{L}_{0}\mathsf{P} & = & \mathsf{L}_{0}\mathsf{P},\label{eq:Ident1}\\
\mathsf{P}\mathsf{L}_{V}\ket{n}\bra{n} & = & 0,\label{eq:Ident2}\\
\mathsf{QL_{0}Q} & = & \mathsf{QL_{0}},\label{eq:Ident3}\end{eqnarray}
which can be proven directly. To further reduce the above expression,
we evaluate the action of $\mathsf{L}_{0}$ and $\mathsf{L}_{V}$
on the electron spin operator $S_{-}$:\begin{eqnarray}
\mathsf{L}_{V}S_{-} & = & -\frac{1}{2}\mathsf{L}_{X}^{+}S_{-},\\
\mathsf{L}_{0}S_{-} & = & \left(-\frac{1}{2}\mathsf{L}_{\omega}^{+}+\mathsf{L}_{D}\right)S_{-},\end{eqnarray}
 where \begin{eqnarray}
\mathsf{L}_{X}^{+}O & = & \commute{X}{O}_{+},\\
L_{\omega}^{+}O & = & \commute{\omega}{O}_{+},\\
L_{D}O & = & \commute{D}{O},\end{eqnarray}
and here we denote anticommutators with a `+' subscript: $\commute{A}{B}_{+}=AB+BA$.
This leads to\begin{multline}
\Sigma^{(2)}(s)=-\frac{i}{4}\mathrm{Tr}_{I}\left[\left(1+\frac{i}{2}\mathsf{L}_{\omega}^{+}\mathsf{Q}\frac{1}{s-\frac{i}{2}\mathsf{L}_{\omega}^{+}}\right)\right.\label{eq:SelfEnergyBornApproxStep1}\\
\left.\times\mathsf{L}_{X}^{+}\frac{1}{s+i\left(\mathsf{L}_{D}-\frac{1}{2}\mathsf{L}_{\omega}^{+}\right)}\mathsf{L}_{X}^{+}\ket{n}\bra{n}\right].\end{multline}
Now, noting that\begin{eqnarray}
\mathsf{Q}\ket{n}\bra{n} & = & 0,\\
\mathsf{Q}\ket{k}\bra{k} & = & \ket{k}\bra{k}-\ket{n}\bra{n},\end{eqnarray}
 we can evaluate Eq. (\ref{eq:SelfEnergyBornApproxStep1}) directly,
giving\begin{multline}
\Sigma^{(2)}(s+i\omega_{n})=-\frac{i}{2}\sum_{k}\left|X_{kn}\right|^{2}\left(\frac{s+\frac{i}{2}\delta\omega_{nk}}{s+i\delta\omega_{nk}}\right)\\
\times\left(\frac{1}{s+i\left(\delta D_{kn}+\frac{1}{2}\delta\omega_{nk}\right)}+\frac{1}{s-i\left(\delta D_{kn}-\frac{1}{2}\delta\omega_{nk}\right)}\right),\label{eq:SelfEnergyBornApproxStep2}\end{multline}
where $\delta D_{kn}=D_{k}-D_{n}$, $\delta\omega_{nk}=\omega_{n}-\omega_{k}$,
and $\omega_{k}$, $D_{k}$ are the eigenvalues associated with eigenstate
$\ket{k}$: $\omega\ket{k}=\omega_{k}\ket{k}$, $D\ket{k}=D_{k}\ket{k}$.
Additionally, we have denoted $X_{kn}=\bra{k}X\ket{n}$. 

From Eqs. (\ref{eq:XDotEquation}), (\ref{eq:SigmaTilde}), and (\ref{eq:LaplaceTransform}),
the electron-spin decoherence rate within a Born-Markov approximation
will now be given by\begin{equation}
\frac{1}{T_{2}}=-\mathrm{Im}\Sigma^{(2)}(s=i(\omega_{n}+\Delta\omega)+0^{+}),\label{eq:BornMarkovDecoherenceRate}\end{equation}
where $0^{+}$ denotes a positive infinitesimal. Our goal here is
to find the leading-order dependence of $1/T_{2}$ on $1/b$ for large
Zeeman splitting: $b>A$. We therefore set $\Delta\omega=-\mathrm{Re}\Sigma^{(2)}(s=i(\omega_{n}+\Delta\omega)+0^{+})\sim\mathcal{O}\left(\frac{A}{N}\left(\frac{A}{b}\right)^{2}\right)\approx0$,
since this term will lead to higher-order corrections in $1/b$ within
the perturbative regime. Additionally, noting that the matrix element
$X_{kn}$ induces a flip-flop for spins at two sites $k_{1,2}$, we
find $\left|\delta D_{kn}\right|=\left|b\left(\gamma_{k_{1}}-\gamma_{k_{2}}\right)\right|$
and $\left|\delta\omega_{kn}\right|=\left|A_{k_{1}}-A_{k_{2}}\right|$.
In the case of a homonuclear system $\gamma_{k_{1}}=\gamma_{k_{2}}$,
we can set $\delta D_{kn}=0$ in Eq. (\ref{eq:SelfEnergyBornApproxStep2}).
Otherwise, in a sufficiently large magnetic field $\left|b\left(\gamma_{k_{1}}-\gamma_{k_{2}}\right)\right|>\left|A_{k_{1}}-A_{k_{2}}\right|$,
we find a negligible contribution to the decoherence rate for terms
from two different isotopic species (where $\gamma_{k_{1}}\ne\gamma_{k_{2}}$),
i.e., heteronuclear flip-flops no longer conserve energy, although
homonuclear fip-flops (for which $\gamma_{k_{1}}=\gamma_{k_{2}}$)
will still occur. Restricting the sum to homonuclear flip-flops and
setting $\delta D_{nk}=0$ in this regime gives\begin{equation}
\Sigma^{(2)}(s+i\omega_{n})=-i\sum_{j}\sum_{k}\left|X_{kn}^{j}\right|^{2}\frac{1}{s+i\delta\omega_{nk}},\label{eq:SigmaHighField}\end{equation}
where $X_{kn}^{j}=\bra{k}X^{j}\ket{n}$ and $X^{j}$ is restricted
to run over flip-flops between nuclei of the common species $j$ at
sites denoted by the indices $k_{j},\, l_{j}$:\begin{equation}
X^{j}=\frac{1}{2}\sum_{k_{j}\ne l_{j}}\frac{A_{k_{j}}^{j}A_{l_{j}}^{j}}{\omega}I_{k_{j}}^{-}I_{l_{j}}^{+}.\label{eq:XjDefinition}\end{equation}
Inserting Eq. (\ref{eq:SigmaHighField}) for a homonuclear system
(one isotopic species $j$) into Eq. (\ref{eq:BornMarkovDecoherenceRate})
and inverting the Laplace transform leads directly to Eq. (10) of
the main text.

\section{\label{sec:Decoherence-rate}Decoherence rate}

Applying Eq. (\ref{eq:BornMarkovDecoherenceRate}) (setting $\Delta\omega\approx0$)
with Eq. (\ref{eq:SigmaHighField}) gives the rate\begin{equation}
\frac{1}{T_{2}}=\pi\sum_{j}\sum_{k}\left|X_{kn}^{j}\right|^{2}\delta\left(\delta\omega_{kn}\right),\label{eq:DecRateGoldenRule}\end{equation}
which can be found directly from the formula\begin{equation}
\frac{1}{x\pm i0^{+}}=\mathcal{P}\frac{1}{x}\mp i\pi\delta(x),\end{equation}
 where $\mathcal{P}$ indicates that the principle value should be
taken in any integral over $x$. Rewriting Eq. (\ref{eq:DecRateGoldenRule})
using the definition of $X^{j}$ given in Eq. (\ref{eq:XjDefinition}):\begin{multline}
\frac{1}{T_{2}}=\frac{\pi}{4}\label{eq:DecRateGoldenRule2}\\
\times\sum_{j}\sum_{k_{j}\ne l_{j}}\frac{c_{-}^{jk_{j}}c_{+}^{jl_{j}}}{\omega_{k}\omega_{n}}\left(A_{k_{j}}^{j}\right)^{2}\left(A_{l_{j}}^{j}\right)^{2}\delta\left(A_{k_{j}}^{j}-A_{l_{j}}^{j}\right),\end{multline}
where $k_{j}$ and $l_{j}$ are restricted to run over sites occupied
by isotopic species $j$. The coefficients $c_{\pm}^{jk_{j}}$ give
the expectation value of the operator $I_{k_{j}}^{\mp}I_{k_{j}}^{\pm}$
with respect to the initial state:\begin{eqnarray}
c_{\pm}^{jk_{j}} & = & \bra{n}I_{k_{j}}^{\mp}I_{k_{j}}^{\pm}\ket{n},\\
 & = & I^{j}(I^{j}+1)-\bra{n}I_{k_{j}}^{z}(I_{k_{j}}^{z}\pm1)\ket{n}.\end{eqnarray}
With small corrections of order $A/Nb\ll1$, we can replace $\omega_{k}\simeq\omega_{n}$
in the denominator of Eq. (\ref{eq:DecRateGoldenRule2}). If the various
nuclear isotopes are uniformly distributed with isotopic concentrations
$\nu_{j}$, we allow the sum over $k_{j},\, l_{j}$ to extend over
all sites $k,l$ at the expense of a weight factor $\nu_{j}$ for
each index:\begin{equation}
\sum_{k_{j}\ne l_{j}}\approx\nu_{j}^{2}\sum_{k\ne l}.\end{equation}
Additionally, we assume that the system is uniformly polarized on
the scale of variation of the hyperfine coupling constants so that
the coefficients $c_{\pm}^{jk}$ can be replaced by average values
$c_{\pm}^{j}=\left<\left<c_{\pm}^{jk}\right>\right>$ (double angle
brackets indicate an average over all sites) and taken out of the
sum. Finally, we change the sums over sites to a double integral using
the prescription and coupling constants described in Appendix \ref{sec:Continuum-limit},
neglecting the small $\mathcal{O}\left(1/N\right)$ correction due
to the requirement $k\ne l$:\begin{equation}
\sum_{k\ne l}\to\int_{0}^{\infty}dk\int_{0}^{\infty}dl.\end{equation}
 These approximations give\begin{multline}
\frac{1}{T_{2}}=\frac{\pi}{4\omega_{n}^{2}}\\
\times\sum_{j}\nu_{j}^{2}c_{-}^{j}c_{+}^{j}\int_{0}^{\infty}dk\int_{0}^{\infty}dl\left(A_{k}^{j}\right)^{2}\left(A_{l}^{j}\right)^{2}\delta\left(A_{k}^{j}-A_{l}^{j}\right).\end{multline}
Inserting the coupling constants defined by Eq. (\ref{eq:AkContLimit})
and evaluating the integrals gives\begin{equation}
\frac{1}{T_{2}}=\frac{\pi}{4}f\left(\frac{d}{q}\right)\sum_{j}\nu_{j}^{2}c_{-}^{j}c_{+}^{j}\frac{A^{j}}{N}\left(\frac{A^{j}}{\omega_{n}}\right)^{2},\label{eq:GeneralRate}\end{equation}
with the geometrical factor $f(d/q)$ given by Eq. (13) of the main
text. Eq. (\ref{eq:GeneralRate}) reduces to Eqs. (12), (14), and
(15) of the main text in the special cases discussed there.

\bibliographystyle{apsrev}
\bibliography{spinecho}

\end{document}